\documentclass[nocopyrightspace]{sigplanconf}

\usepackage[all]{xy}
\usepackage[dvips]{graphicx}
\usepackage{paper}

\usepackage[bookmarksopen=true]{hyperref}

\usepackage{paralist}
\conferenceinfo{PLDI'13,} {June 16--19, 2013, Seattle, WA, USA.}
\CopyrightYear{2013}
\copyrightdata{978-1-4503-2014-6/13/06}

\title{Quipper: A Scalable Quantum Programming Language}

\authorinfo{\vspace{-0.6cm} Alexander S. Green}
           {Dalhousie University}
           {agreen@mathstat.dal.ca}
\authorinfo{\vspace{-0.6cm} Peter LeFanu Lumsdaine}
           {Institute of Advanced Studies}
           {p.l.lumsdaine@gmail.com}
\authorinfo{\vspace{-0.6cm} Neil J. Ross}
           {Dalhousie University}
           {Neil.JR.Ross@Dal.Ca}
\authorinfo{\vspace{-0.5cm} Peter Selinger}
           {Dalhousie University}
           {selinger@mathstat.dal.ca \\ \vspace{-1cm}}
\authorinfo{\vspace{-0.5cm} Beno\^{i}t Valiron}
           {University of Pennsylvania}
           {benoit.valiron@monoidal.net \\ \vspace{-1cm}}

\begin{document}

\maketitle

\begin{abstract}
  
  The field of quantum algorithms is vibrant. Still, there is
  currently a lack of programming languages for describing quantum
  computation on a practical scale, i.e., not just at the level of toy
  problems. We address this issue by introducing Quipper, a scalable,
  expressive, functional, higher-order quantum programming language.
  Quipper has been used to program a diverse set of non-trivial
  quantum algorithms, and can generate quantum gate
  representations using trillions of gates. It is geared towards a
  model of computation that uses a classical computer to control a
  quantum device, but is not dependent on any particular model of
  quantum hardware. Quipper has proven effective and easy to use, and
  opens the door towards using formal methods to analyze quantum
  algorithms.
\end{abstract}


\section{Introduction}

The earliest computers, such as the ENIAC and EDVAC, were both rare
and difficult to program. The difficulty stemmed in part from the need
to express algorithms in a vocabulary suited to the particular
hardware, ranging from function tables for the ENIAC to more
conventional arithmetic and movement operations for later
machines. The introduction of symbolic programming languages such as
FORTRAN (for ``FORmula TRANslator'') solved a major difficulty for the
next generation of computing devices, by enabling the specification of
algorithms in a form more suitable for human understanding, and then
translating this specification into a form executable by the
machine. Thus, programming languages assumed the important role of
bridging a semantic gap between the human and the computing
device. This was achieved, among other things, by two important
principles: high-level abstractions and automated bookkeeping.

Quantum computation, which was envisioned in the later part of the
20th century, is a computational paradigm based on the laws of quantum
physics. It has been amply demonstrated in the literature that quantum
computing can, in theory, outperform classical computing for certain
classes of computational problems. The design of new quantum
algorithms is a vibrant area, as witnessed by the quantum algorithm
``zoo'' of S.  Jordan~\cite{jordan-qzoo}, which references 45
algorithms and 160 papers, with no less than 14 written in 2011 and
2012.

Although quantum computing is not yet ready to move from theory to
practice, it is nevertheless possible to make informed guesses of what
form an eventual quantum computer may take, or more importantly for
programming language design, of the {\em interface} by which one may
interact with such a quantum computer. It seems wise, then, to apply
the lessons learned from programming classical computing to the
emerging quantum computing capabilities.

This paper is a stepping stone towards meeting this challenge. We
approach quantum computation from a programmer's perspective: how
should one design a programming language that can implement real-world
quantum algorithms in an efficient, legible and maintainable way?  We
introduce Quipper, a declarative language with a monadic operational
semantics that is succinct, expressive, and scalable, with a sound
theoretical foundation.

When we speak of Quipper being ``scalable'', we mean that it goes well
beyond toy algorithms and mere proofs of concept.  Many actual quantum
algorithms in the literature are orders of magnitude more complex than
what could be realistically implemented in previously existing quantum
programming languages. We put Quipper to the test by implementing
seven non-trivial quantum algorithms from the literature:
\begin{compactitem}
\item Binary Welded Tree (BWT). To find a labeled node in a
  graph~\cite{BWT}.
\item Boolean Formula (BF). To evaluate a NAND formula~\cite{BF}. The
  version of this algorithm implemented in Quipper computes a winning 
  strategy for the game of Hex.
\item Class Number (CL). To approximate the class group of a real
  quadratic number field~\cite{CN}.
\item Ground State Estimation (GSE). To compute the ground state
  energy level of a particular molecule~\cite{GSE}.
\item Quantum Linear Systems (QLS). To solve a linear system of
  equations~\cite{LS}.
\item Unique Shortest Vector (USV). To choose the shortest vector
  among a given set~\cite{SV}.
\item Triangle Finding (TF). To exhibit a triangle inside a dense
  graph~\cite{TF}.
\end{compactitem}
These algorithms were chosen by IARPA, in the context of its QCS
program {\cite{BAA}}, to provide a reasonably representative 
cross-section of current algorithms. They make use of a wide variety
of quantum primitives, such as amplitude amplification, quantum walks,
the quantum Fourier transform, and quantum simulation. Several of the
algorithms also require the implementation of complex classical
oracles. The starting point for each of our algorithm implementations
was a detailed description of the algorithm provided by IARPA.

\paragraph{Related work.}

Many formalisms for programming quantum computers have been developed
in the last few decades.  Some of them, such as the quantum Turing
machine~\cite{Deutsch-1985} or the quantum lambda calculus of van
Tonder~\cite{Tonder-2004}, are mainly theoretical tools for exploring
particular aspects of quantum computation, and are not designed with
practical quantum programming in mind.

There are many recent proposals for quantum programming
languages~\cite{Gay-2006}. Of these, we pinpoint three languages that
represent important milestones and can be regarded as predecessors of
Quipper.

In the realm of imperative programming languages, arguably the oldest
``concrete'' quantum programming language is \"Omer's
QCL~\cite{Omer-2000}. Defined as a C-style language, QCL comes with
many interesting features, collectively dubbed {\em structured quantum
  programming}. This provides a relatively natural way of writing
simple quantum algorithms.  One of QCL's innovations was the
separation of functions into separate syntactic classes, based on
their operational behavior; thus, QCL distinguishes classical
procedures, which are unconstrained; ``quantum functions'', which are
restricted to define unitary operations; and ``pseudo-classical''
operators, which are intended to implement oracles, featuring
``quantum tests'' and automatic uncomputation of ancillas. QCL lacks
high-level quantum data types, and does not have a well-defined
semantics, complicating the analysis of programs. Finally, since the
language was designed with simulation in mind, many of its useful
programming features incur a strong computational overhead.  In
spite of these drawbacks, QCL is a milestone in the development of
quantum programming languages. We include a very brief comparison
between circuits generated by Quipper and QCL in
Section~\ref{sec:qcl}.

More recently, there have been two proposals for functional quantum
programming languages that can be regarded as precursors of
Quipper. Selinger and Valiron's quantum lambda calculus is an ML-style
language with strong static type
checking~\cite{Selinger-Valiron-2006,Selinger-Valiron-2009}. It is
designed to run on Knill's QRAM model {\cite{Knill-1996}}, but lacks
high-level facilities for circuit construction and manipulation. The
quantum IO Monad of Green and Altenkirch~\cite{Altenkirch-Green-2009}
is, like Quipper, embedded in Haskell, provides extensible quantum
data types, and comes with a consistent operational
semantics. However, it uses a much simpler circuit model and lacks
many of Quipper's advanced programming features.

\paragraph{Outline of the paper.}

In Section~\ref{sec:qc}, we briefly present quantum computation,
focusing particularly on the interface by which software would
interact with a quantum device. Section~\ref{sec:techniques} covers
some of the main techniques that are used to describe quantum
algorithms and hopefully makes the case for a quantum programming
language. In Section~\ref{sec:quipper}, we introduce Quipper.
Section~\ref{sec:triangle} discusses our implementation of the
Triangle Finding algorithm, and Section~\ref{sec:qcl} contains a very
brief comparison between Quipper and QCL. We summarize our conclusions
at the end.

\section{Quantum computation}
\label{sec:qc}

We very briefly summarize some basic notions from quantum computation,
primarily to provide hints on how a quantum programming language might
interact with a quantum computer. One cannot really do this subject
justice in such a limited space. For a much more thorough introduction
to quantum computing, see e.g.  {\cite{Nielsen-Chuang-2002}}.

In quantum computation, the storage and manipulation of data is
governed by the law of quantum physics. We will here be concerned with
{\em idealized} quantum computation, i.e., we ignore the effects of
physical imprecisions, decoherence, etc. We will describe an idealized
quantum device in terms of its {\em state} and {\em operations}.

The state of a quantum system is given by a normalized vector in a
Hilbert space. The smallest unit of information in quantum computing
is the {\em quantum bit} or {\em qubit}; the state of one qubit is a
complex linear combination of two basis vectors $\ket{0}$ and
$\ket{1}$. Similarly, the state of two qubits is given as a linear
combination of four basis vectors $\s{\ket{00}, \ket{01}, \ket{10},
  \ket{11}}$, and more generally, the state of $n$ qubits is a linear
combination of $2^n$ basis vectors. The available operations are {\em
  unitary transformations}, which allow the state to be transformed
along a user-specified unitary map; and {\em measurements}, which are
the only way to extract classical information from a quantum state. We
usually assume that each quantum device has some built-in set of
elementary unitary transformations, called {\em gates}. Measurement
has a probabilistic behavior: for example, when measuring a qubit in
state $\alpha\ket{0} + \beta\ket{1}$, the result will be $0$ with
probability $|\alpha|^2$ and $1$ with probability $|\beta|^2$, and
subsequently the state of the qubit will have been changed to
$\ket{0}$ or $\ket{1}$, respectively. There is an analogous rule for
measuring, say, one of several qubits in a multi-qubit state. 

\subsection{Interacting with a quantum device}\label{ssec:interacting}

We can now describe the operation of an idealized quantum device known
as Knill's QRAM model for quantum computation~\cite{Knill-1996}. In
this model, we think of a quantum computer as a specialized device
that is attached to and controlled by a classical computer, much in
the way of a co-processor. The device holds $n$ individually
addressable qubits, for some fixed $n$. The operation of the quantum
device is controlled by only two kinds of instructions, which can be
interleaved. Instructions
of the first kind are unitary operations. They take the form ``apply
the built-in unitary gate $U$ to qubit $k$'', ``apply the gate $V$ to
qubits $j$ and $k$'', and so on. The quantum device responds with an
acknowledgement that the operation has been performed, but there is no
further information returned. Instructions of the second kind are
measurements. They take the form ``measure qubit $k$''. The quantum
device responds with a measurement result, which is either $0$ or $1$.
One can also add a third kind of instruction called {\em
  initialization}: ``reset qubit $k$ to $0$''. However, this is
derivable from the instructions already mentioned: namely, by first
measuring qubit $k$, and then negating it if and only if the
measurement outcome was $1$.

\subsection{Basic properties}

In the above model of quantum computation, the control flow of an
algorithm is purely classical: tests, loops, etc., are performed on
the classical computer that controls the quantum co-processor. Both
classical and quantum data are first class objects.

Because quantum measurement is a probabilistic operation, classical
probabilistic computation is automatically included as a subset of
quantum computation.

The laws of quantum mechanics imply that quantum information cannot be
duplicated. This is the so-called {\em no-cloning} property of quantum
mechanics. It would not be physically meaningful, for example, to
apply a 2-qubit quantum gate to qubits $k$ and $k$. Quantum
programming languages should ensure that such non-physical operations
cannot occur. This kind of property can either be checked at
compile time or at run time.

\subsection{Hardware independence}

We do not claim that the idealized QRAM model is what an actual
quantum computer will look like. An actual quantum computer might be
far more difficult to control. Because of the relatively short life
span of quantum states in experimental settings, many layers of
quantum error correction and control will likely be required to enable
meaningful quantum computation. Also, realistic quantum hardware may
be highly sensitive to timing constraints, such as the exact timing of
control pulses. So rather than performing one gate or measurement at a
time, as suggested in the QRAM model, it may be more realistic to
assume that a large number of gates will be pre-computed, then
executed in a single batch operation on the quantum device, possibly
measuring all qubits at the end. A sequence of pre-computed gates is
called a {\em quantum circuit}, and this model of quantum computation
is known as the {\em circuit model}.

One operation that is available in the QRAM model, but not in the
circuit model, is the ability to change the sequence of quantum gates
in response to the results of previous measurements. This restriction
can be overcome by augmenting the circuit model with the ability to
preserve some of the unmeasured qubits in some kind of long-term
storage between successive circuit executions.

From the point of view of programming language design, the particular
choice of physical quantum architecture should not be of much
consequence. The purpose of a high-level programming language is
precisely to abstract from such hardware specific details, and to
present the user with the illusion of a uniform idealized
computational model.

\section{Techniques used in quantum algorithms}
\label{sec:techniques}

While every quantum algorithm can be ultimately specified as a
sequence of gates and measurements, this is rarely how quantum
algorithms are actually described in the literature. Rather, they are
often described at a high level, for example in the style of: ``Take
the following function, which can obviously be implemented by a
boolean circuit of polynomial size. Translate this to a reversible
quantum circuit in the standard way. Apply $m$ steps of amplitude
amplification, then copy the result to a scratch register and
uncompute''. We believe that a good quantum programming language
should be flexible enough to allow quantum algorithms to be expressed
at a level of abstraction, high or low, that is as close as possible
to the intent of the algorithm's human designer, while filling in
enough details to be unambiguous. For this reason, prior to introducing
Quipper's high-level programming features in the next section, let us
briefly review some of the techniques that are commonly used in the
design of quantum algorithms.

\subsection{Quantum primitives}

Most quantum algorithms make use of one or more of a few well-known
primitive building blocks. The {\em quantum Fourier transform} is
a unitary change of basis analogous to the classical Fourier
transform, and is used in many quantum algorithms, for example to find
the period of a periodic function. 
{\em Amplitude amplification} (also known as {\em Grover's search}) is
used to increase the amplitude of certain basis states in a
superposition, while decreasing others.
{\em Quantum walks} can be described as the quantum counterpart to random
walks. Due to quantum interference, some paths in the walk may cancel
out (or at least, appear with decreased probability). In some
situations, it is possible to outperform the success probability of a
similar strategy that would have used a classical random walk.
{\em Phase estimation} is a technique for estimating eigenvalues of a
unitary operator.
{\em State distillation} is a method by which one starts with a large
noisy set of quantum states, and gradually narrows them down to a smaller
cleaner set of states with desirable properties.

The above primitives are often at the heart of what makes a quantum
algorithm potentially outperform its classical counterpart. But they
are more than just off-the-shelf functions that can be directly used
on a classical data structure, and they are typically combined in
non-trivial ways.

\subsection{Oracles}\label{ssec:oracles}

Another important part of many quantum algorithms is the description
of an {\em oracle}. An oracle is usually given by a classical function
$f:\Bool^n\to\Bool^m$, describing some aspect of the input to the
algorithm, such as the edges of a graph, the winning positions of a
game, arithmetic or number-theoretic functions, and so forth. To be
useable in a quantum computation, the oracle must be made
reversible. This can be done by lifting the function, such that
$\hat{f}:\Bool^{n+m}\to\Bool^{n+m}$ is defined as
$\hat{f}(x,y)=(x,y\oplus f(x))$. The reversible boolean function
$\hat{f}$ can then be lifted into a unitary map working on quantum
bits. Often, in the literature, the description of oracles is both
low-level and high-level. It is low-level in the sense that, despite
the fact that the oracle manipulates non-trivial data types
(e.g., integers, real numbers, edges of a graph, etc.), the algorithm
goes into detail about how to implement these in terms of quantum
registers. But it is also high-level, in the sense that the details of
how the oracle performs its operations are often only sketched.

\subsection{Circuit families}

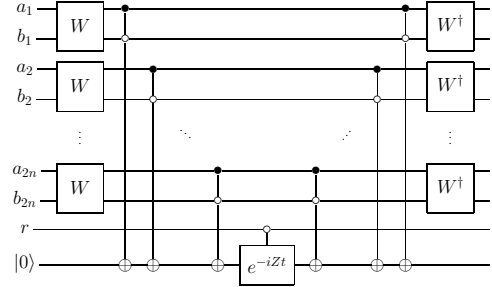
\begin{figure}
  \newlength{\size}
  \setlength{\size}{1.3em}
  \def\r#1#2{\raisebox{#1}{#2}}\def\h#1{\hspace{#1}}
  \def\myvdots{\r{.4\size}{\vdots}}
  \def\c{*-={\mbox{\Large$\circ$}}}\def\b{*-={\mbox{\Large$\bullet$}}}\def\vd{*+={\myvdots}}\def\cd{\cdots}
  \def\wtop{\r{.4\size}{\h{2.5\size}}}
  \def\wbot{\r{-.4\size}{\h{2.5\size}}}
  \def\aa{\ar@{-}}
  \def\op{*-={\mbox{\Large$\oplus$}}}
  \def\Ddots{{{.}\h{.2pt}{\r{2.8pt}{.}}\h{.2pt}{\r{5.8pt}{.}}}}
  \def\makew#1#2{\save
    "#1,2";"#2,2"**@{}?<(.4)*{\mbox{\Large$W$}};"#1,2"."#2,2"*\frm{-};
    "#1,12";"#2,12"**@{}?<(.3)*{\mbox{\Large$W^{\dagger}$}};"#1,12"."#2,12"*\frm{-};
    \restore}
  \centering
  \scalebox{.5}{$\xymatrix@C=\size@R=.5\size{
      \mbox{\Large$a_1$} \aa[r] &\wtop \aa[r]&
      \b\aa[rrrrrrrr]\aa[d]&&&&&&
      &&\b\aa[r]\aa[d]&\wtop\aa[r]&
      \\
      \mbox{\Large$b_1$} \aa[r] &\wbot \aa[r]&
      \c\aa[rrrrrrrr]\aa[ddddddd]&&&&&&
      &&\c\aa[r]\aa[ddddddd]&\wbot\aa[r]&
      \makew12
      \\
      \mbox{\Large$a_2$} \aa[r] &\wtop \aa[rr]&&
      \b\aa[rrrrrr]\aa[d]&&&&&
      &\b\aa[rr]\aa[d]&&\wtop\aa[r]&
      \\
      \mbox{\Large$b_2$} \aa[r] &\wbot \aa[rr]&&
      \c\aa[rrrrrr]\aa[ddddd]&&&&&
      &\c\aa[rr]\aa[ddddd]&&\wbot\aa[r]&
      \makew34
      \\
      &\vd & & &\ddots & & & &
      \Ddots& & &\vd&
      \\
      \mbox{\Large$a_{2n}$} \aa[r] &\wtop \aa[rrrr]&&
      &&\b\aa[rr]\aa[d]&&
      \b\aa[rrrr]\aa[d]&&&&\wtop\aa[r]&
      \\
      \mbox{\Large$b_{2n}$} \aa[r] &\wbot \aa[rrrr]&&
      &&\c\aa[rr]\aa[dd]&&
      \c\aa[rrrr]\aa[dd]&&&&\wbot\aa[r]&
      \makew67
      \\
      \mbox{\Large$r$}
      \aa[rrrrrr]&&&&&&\c\aa[rrrrrr]\aa[d]
      &&&&&&&
      \\
      \mbox{\Large$|0\rangle$}\aa[rr]&&\op\aa[r]&\op\aa[rr]&
      &\op\aa[r]&*+<1\size,1.5\size>[F]{\mbox{\Large$e^{-iZt}$}}\aa[r]&
      \op\aa[rr]&&\op\aa[r]&
      \op\aa[rr]&&
    }$}
  \caption{Example of a quantum circuit. Circuits are read left to
    right, with horizontal lines representing wires, boxes
    representing quantum gates, and vertical wires representing
    controls on a gate.}
  \label{fig:bwt}
\end{figure}

At a low level, quantum algorithms take the form of a (potentially
very long) sequence of unitary gates with occasional
measurements. Such a sequence of operators is called a {\em quantum
  circuit} and is customarily described in diagrammatic form. An
example of such a diagram, showing a diffusion step from the Binary
Welded Tree algorithm~\cite{BWT}, is shown in Figure~\ref{fig:bwt}.
However, such diagrams are not in and by themselves good descriptions
of quantum algorithms. The reason is that most quantum algorithms also
depend on {\em parameters}, such as the number $n$ in
Figure~\ref{fig:bwt}, and thus a quantum algorithm really describes a
{\em family} of circuits, which cannot be captured in a single
diagram.  Quipper permits a formal and precise description of such
parameterized circuit families.

\subsection{Circuit manipulation}

Although ultimately, a quantum algorithm comes down to a sequence of
elementary gates and measurements, many quantum algorithms are more
naturally described in terms of manipulations at the level of entire
sub-circuits, rather than individual gates. Examples of such
operations are:
\begin{compactitem}
\item reversing;
\item iteration (e.g., Trotterization; amplitude amplification);
\item automatic synthesis of classical circuits (e.g., oracles) and
  ancilla management (i.e., initialization and recollection of auxiliary
  quantum bits);
\item circuit transformations (e.g., replacing one elementary gate set
  by another);
\item whole-circuit optimizations.
\end{compactitem}

\subsection{Classical processing}

To be useful, a complete quantum program must ultimately produce
a classical answer to a classical question. In particular, any
parameters to the algorithm are classical, as are the final
outputs. Therefore, most quantum algorithms use some amount of
classical pre- and post-processing. Typically, the algorithm consists
of the description of a parameterized quantum circuit, followed by a
final measurement. 

In some algorithms, such as the Triangle Finding algorithm, the
probabilistic measurement result can then be classically checked to
see if a useful answer has been found, and if not, the whole procedure
is repeated, possibly for a different set of parameters.  In some
algorithms, such as the Binary Welded Tree algorithm, the validity of
a potential solution cannot be efficiently verified, and a statistical
argument is used to determine how many times the algorithm should be
repeated until the correct answer is found with the desired
probability. A third class of algorithms, such as the Unique Shortest
Vector algorithm, requires a more subtle interleaving of quantum and
classical operations, whereby only a subset of the qubits are
measured, and the quantum memory cannot be reset between each quantum
circuit invocation. In the paradigm of quantum circuits, this amounts
to saying that the circuit is constructed on-the-fly, where later
pieces depend on the value of former intermediate measurements.  This
is typically the case for algorithms that incorporate state
distillation.

We learn from this that a usable quantum programming language should
also incorporate a general-purpose classical programming language, in
which classical pre-, post-, and intermediate computations can be
specified. It is desirable that the integration between the classical
and quantum parts of the language is as seamless as possible, but that
a clear distinction still exists.

\section{Our proposal: Quipper}
\label{sec:quipper}

We introduce Quipper, an embedded functional programming language for
quantum computation. Quipper is intended to offer a unified
general-purpose programming framework for quantum computation. It
provides, among other things, a notation for quantum circuits, a
notation for quantum algorithms, and a notation for circuit
transformations. 

Quipper was designed with correctness, scalability and usability in
mind. It was originally developed in the context of IARPA's Quantum
Computer Science program {\cite{BAA}}. We have demonstrated Quipper's
viability by implementing seven non-trivial quantum algorithms from
the literature {\cite{BWT,BF,CN,GSE,LS,SV,TF}}, as selected by IARPA
{\cite{BAA}}. In this section, we describe some of the basic features
of Quipper's design.

\subsection{Quipper is an embedded language}

We implemented Quipper as an embedded language, with Haskell as the
host language. Therefore, Quipper can be seen as a collection of data
types, combinators, and a library of functions within Haskell,
together with an {\em idiom}, i.e., a preferred style of writing
embedded programs. See {\cite[Sec.~1.3]{Claessen-2001}} for a general
discussion of the advantages and disadvantages of embedded languages
in programming language design.

We chose Haskell as the host language because Quipper contains many
higher-order and overloaded operators, whose implementation makes
heavy use of advanced features of Haskell's type system, including
several GHC extensions. Both Haskell and Quipper are strongly-typed
functional programming languages, and therefore they are a relatively
good fit for each other. Of course, there are some trade-offs. In
particular, Haskell lacks two features that would be useful for
Quipper: {\em linear types} and {\em dependent types}.  Therefore,
certain properties of quantum programs that could be checked at
compile time by a linear or dependent type system must currently be
checked at run-time. For this reason, a future implementation of
Quipper may be equipped with a stand-alone compiler, or at least a
custom type-checker.

\subsection{Quipper's extended circuit model}
\label{subsection:circuit-model}

The quantum circuit model, as usually presented (see e.g. 
{\cite[Sec.~4]{Nielsen-Chuang-2002}}), is only concerned
with unitary gates and circuits. While this is theoretically
sufficient, we found it to be a cumbersome restriction in
practice. Quipper natively supports a larger class of circuits that
also includes:
\begin{compactitem}
\item Explicit qubit initialization and termination. This is useful,
  among other things, for accurately representing the scope of
  ancillas.
\item Measurements, classical bits, classical gates, and
  classically-controlled quantum gates.
\end{compactitem}

\subsubsection{Ancillas and scope}

Many quantum algorithms require ancillas, i.e., ``scratch space''
qubits whose state is (say) $\ket{0}$ outside of certain well-defined
regions where the ancilla is being ``used''. In settings where all
gates must be unitary, ancillas are usually treated as additional
global inputs and outputs to the algorithm, which are assumed to be in
state $\ket{0}$ at the start of the algorithm, and which the algorithm
is expected to reset to $\ket{0}$ after each ``use''. The following
image shows a circuit with two ancillas, and the regions where the
ancilla is in state $\ket{0}$ are highlighted:

\begin{center}
  \includegraphics[width=.9\columnwidth]{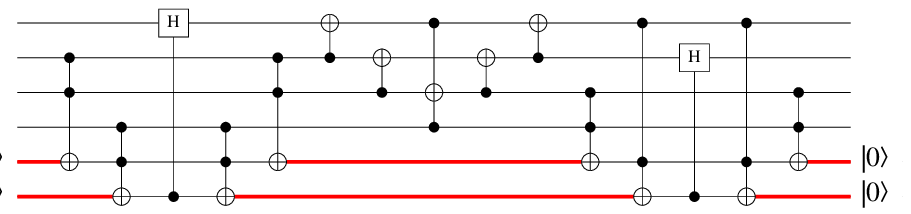}
\end{center}

We refer to the regions where an ancilla may potentially be used as
the {\em scope} of the ancilla.  For a compiler of quantum programming
languages, there are many potential benefits to tracking the scope of
ancillas explicitly. For example, it would be wasteful for error
correction to be applied to an ancilla while it is known to be unused
(and therefore disentangled from the rest of the
computation). Moreover, if an algorithm temporarily requires two
ancillas at some point in time, and then again two ancillas at some
later time, it does not actually matter whether the two later ancillas
are ``equal'' to the earlier ancillas, whether they are swapped, or
whether they are different ancillas altogether. For example, the
following circuit is equivalent to the one above:

\begin{center}
  \includegraphics[width=.9\columnwidth]{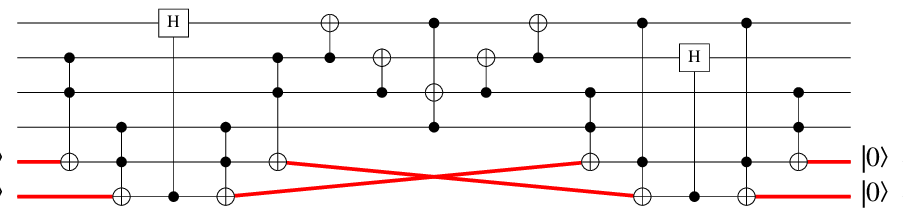}
\end{center}

The problem of which particular ancillas to use from a ``pool'' of
ancillas is analogous to the classical problem of register allocation,
and is best left to a late compiler phase that is aware of the layout
of physical qubits.

In Quipper's circuit model, we use the notation ``$0\Vdash$'' to
denote the allocation of a new qubit initialized to state
$\ket{0}$. Dually, we use the notation ``$\dashV 0$'' to denote the
deallocation of a qubit that is {\em asserted} to be in stated
$\ket{0}$.  Here is the same circuit as above, represented with
explicitly scoped ancillas:

\begin{center}
  \includegraphics[width=.9\columnwidth]{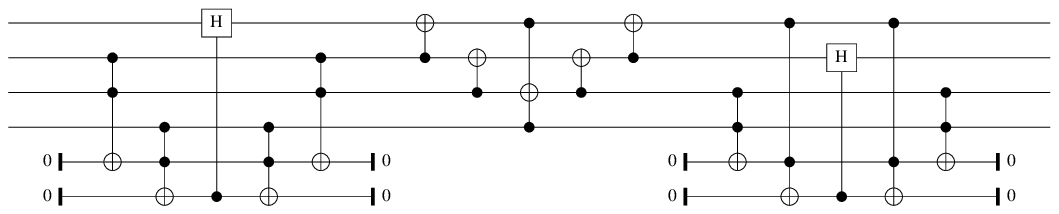}
\end{center}

Keeping track of ancilla scopes also has an additional possible
advantage. In certain physical machine models, such as photonics, it
is generally better to work with ``fresh'' photons than with photons
that have been in a holding loop. This is because photons have a
relatively high dissipation rate. Scoped ancillas were used
extensively in our seven algorithm implementations.

\subsubsection{Assertive termination}

As explained above, the gate $\dashV 0$ terminates (or deallocates) a
qubit while asserting that it is in state $\ket{0}$. We call this an
{\em assertive termination}, to distinguish it from the ordinary
termination, denoted $\dashV$, which simply drops the qubit (therefore
resulting in a possibly mixed state).

The concept of assertive qubit termination warrants some further
thoughts. The first thing to note is that it is the {\em programmer},
and not the compiler, who is asserting that the qubit is in state
$\ket{0}$ before being terminated. In general, the correctness of such
an assertion depends on intricacies of the particular algorithm, and
is not something that the compiler can verify automatically. It is
therefore the programmer's responsibility to ensure that only correct
assertions are made. The compiler is free to rely on these assertions,
for example by applying optimizations that are only correct if the
assertions are valid.

The second thing to note is that circuits containing qubit
initializations and assertive terminations can never result in a mixed
state, and are, in a suitable sense, unitary and reversible. More
precisely, where assertive qubit terminations are used in a circuit,
they determine a certain {\em subspace} of its domain: namely, the
subspace of those states for which the assertions are true. Dually,
the use of qubit initializations determines a certain subspace of the
co-domain: namely, the subspace of states that are reachable (or
equivalently, in the {\em image} of the circuit). The circuit then
defines a unitary bijection between these two subspaces. In
particular, it follows that a circuit using $n$ input qubits and $n$
output qubits, and using any number of local ancillas, is unitary
(provided, of course, that all termination assertions are correct,
i.e., all ancillas are uncomputed correctly). For this reason, Quipper
will, without complaint, reverse circuits containing qubit
initializations and assertive terminations.

\subsubsection{Mixed classical/quantum circuits}

In the circuit model used by Quipper, classical and quantum data can
co-exist. Classical wires (whose state is a classical bit), classical
gates, and classically-controlled quantum gates can be freely combined
with pure quantum gates. Measurement is a gate that turns a qubit into
a classical bit.  One reason for including these features is the
construction of oracles, which we will discuss in more detail in
Section~\ref{ssec:quipperoracle}.

\subsection{The two run-times}

\subsubsection{Circuit generation and circuit execution}

Because Quipper is (among other things) a circuit description
language, Quipper programs have three distinct phases of execution:
compile time, circuit generation time, and circuit execution time.
We refer to circuit generation time and circuit execution times as the
``two run-times''. The phenomenon of having three distinct phases of
execution is well-known and also occurs, for example, in hardware
description languages (see e.g. {\cite{Claessen-2001}}).

\begin{enumerate}
\item Compile time. Since Quipper is an embedded language, its compile
  time is the same as the Haskell compile time. It takes place on a
  classical computer in an off-line development environment (i.e.,
  before specific algorithm parameters are known). The input to this
  phase is source code and {\em compile time parameters}. The output
  is executable object code.
\item Circuit generation time. This takes place on a classical
  computer in an on-line environment (i.e., when specific algorithm
  parameters are known). The input to this phase is executable object
  code and {\em circuit parameters} (for example, the size of
  registers, problem sizes, the size of time steps, error thresholds,
  etc.). The output is a representation of a quantum circuit.
\item Circuit execution time. This takes place on a physical quantum
  computer in an on-line real-time environment. The input to this
  phase is a quantum circuit, and possibly some {\em circuit inputs}
  (for example, qubits fetched from long-term storage to initialize
  circuit inputs, if supported by the physical device; classical bits
  to be used as classical circuit inputs). The output consists of
  circuit outputs (for example, classical bits that are measurement
  results; qubits to be moved to long-term storage, if supported). 
\end{enumerate}

Many quantum algorithms require an alternation between the second and
third phases
(circuit generation time and circuit execution time). In this model of
execution, the classical controller generates a circuit, sends it to
the physical device for execution, awaits measurement results, then
generates another circuit, and so on. We note that this is the same as
the usual quantum circuit model of computation.  If, moreover, the
physical quantum device has the ability to preserve qubits in
long-term storage between real-time circuit invocations, then one can
support a more general model of computation known in Quipper as {\em
  dynamic lifting}: this allows circuit outputs (for example, the
results of measurements) to be re-used as circuit parameters (to
control the generation of the next part of the circuit).  An example of
such a model of computation is Knill's QRAM model
{\cite{Knill-1996}}. We believe that Quipper's abstract computational
paradigm is general enough to support a variety of such concrete
computational models.

\subsubsection{The parameter/input distinction}\label{sssec:parameter-input}

We use the word ``parameter'' to refer to a value that is known at
circuit generation time, and we use the word ``input'' or ``state'' to
refer to a value that is only known at circuit execution time, i.e.,
the state of a bit or qubit on the physical quantum device, thought of
as a ``wire'' in a circuit. The distinction between inputs and
parameters must be taken seriously and requires special programming
language support. For example, because inputs are not known at circuit
generation time, if one would like to do an if-then-else operation
conditioned on a boolean {\em input}, then one must generate the
circuit for the then-part {\em and} the else-part. On the other hand,
if the if-then-else operation is conditioned on a boolean {\em
  parameter}, then one only needs to generate the circuit for the
then-part {\em or} the else-part, resulting in a smaller circuit.

Because of this distinction between generation-time parameters and
execu\-tion-time inputs, the Quipper language has three basic types for
bits and qubits, instead of the usual two:

\begin{compactitem}
\item {\tt Bool}: a boolean parameter, known at circuit generation
  time;
\item {\tt Bit}: a boolean input, i.e., a boolean wire in a circuit;
\item {\tt Qubit}: a qubit input, i.e., a quantum wire in a circuit.
\end{compactitem}

A {\tt Bool} is a parameter and can be easily converted to a {\tt
  Bit}. The outcomes of quantum measurements are only known at circuit
execution time, and are therefore Bits, not Bools. As mentioned above,
the converse operation, converting a {\tt Bit} to a {\tt Bool}, is
known as {\em dynamic lifting} in Quipper, and is usually an expensive
operation, requiring circuit execution to be suspended while the next
part of the circuit is generated. 

The input/parameter distinction also applies to classical data types
other than booleans; for example, there are integer parameters and
integer inputs. 

Moreover, some data is partly input and partly parameter. For example,
if a quantum function inputs a list of qubits, then the {\em length}
of the list is a parameter (affecting, for example, circuit size),
whereas the actual qubits in the list are inputs. In Quipper
terminology, when a piece of data has both input and parameter
components, the parameter component is called the {\em shape} of the
data.

\subsection{Circuit description language}
\label{subsection:circuit}

One can readily imagine a quantum programming language that operates
by sending gate-by-gate instructions in real time to some physical
quantum device. Indeed, this was the approach taken in 
{\cite{Selinger-Valiron-2006,Selinger-Valiron-2009}}. However, we
found that this approach is not very practical when it comes to
implementing larger-scale quantum algorithms. Quantum algorithms in
the literature are often represented at a relatively high conceptual
level, and many tasks in algorithm construction require manipulations
at the level of entire circuits, rather than individual
gates. Examples of such operations include inversion; iteration;
ancilla management; circuit transformations (e.g., replacing one set
of basic gates by another); and whole-circuit optimization. Another
important use of whole-circuit manipulation is the automatic
generation of reversible circuits from classical code. In our
experience, it is perhaps fair to say that 99 percent of the quantum
programmer's task is constructing and manipulating circuits, and only
1 percent is actually running them.

We therefore designed Quipper with the goal of supporting both
gate-level operations and circuit-level operations in a natural
way. Quipper combines a basic procedural paradigm for writing quantum
functions ``one gate at a time'' with a powerful higher-order paradigm
for whole-circuit manipulations.

\subsubsection{Procedural paradigm}

The basic philosophy of Quipper's procedural paradigm is that qubits
are held in variables and gates are applied to them one at a
time. Subroutines can be used to group gate-level operations together
where the programmer finds it useful. When writing such procedural
code, the programmer may safely pretend --- although this is not
actually true --- that the variables hold actual physical qubits, and
that the specified gates are applied to them in real time. 

Thus, the basic abstraction offered by Quipper is that a quantum
operation is a {\em function} that inputs some quantum data, performs
state changes on it, and then outputs the changed quantum data. This
is encapsulated in a Haskell monad called {\tt Circ}. For example, the
following is a simple quantum function that inputs a pair of quantum
bits, performs some unitary operations (two Hadamard gates and a
controlled not-gate), and outputs the modified pair of quantum
bits. The code is shown on the left, and the generated circuit is
shown on the right.

\noindent\begin{minipage}{\columnwidth}
\begin{mycode}
\begin{verbatim}
mycirc :: Qubit -> Qubit -> Circ (Qubit, Qubit)
mycirc a b = do
  a <- hadamard a
  b <- hadamard b
  (a,b) <- controlled_not a b
  return (a,b)
\end{verbatim}
\end{mycode}

\hspace{2in}\includegraphics[scale=1.3, bb = 0 -10 1 -10]{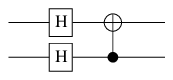}
\end{minipage}

Gates can also be written in ``imperative style'', i.e., the return
value of a gate can be ignored if it consists of the same physical
qubits as the gate's input. For now, this is just a notational
convention, but it could be formalized through the use of a
\emph{linear type system} in a future version of Quipper.

\subsubsection{Block structure}

Quipper provides operators for introducing block structure into
circuits. For example, the operator
\begin{mycode}
\begin{verbatim}
with_controls :: Qubit -> Circ a -> Circ a
\end{verbatim}
\end{mycode}
can be used to let an entire block of gates be controlled by a
qubit. The example also illustrates how subroutines (in this case,
{\tt mycirc} defined above) can be used to build up complex circuits
from simpler ones.

\noindent\begin{minipage}{\columnwidth}
\begin{mycode}
\begin{verbatim}
mycirc2 :: Qubit -> Qubit -> Qubit 
  -> Circ (Qubit, Qubit, Qubit)
mycirc2 a b c = do
  mycirc a b
  with_controls c $ do
    mycirc a b
    mycirc b a
  mycirc a c
  return (a,b,c)
\end{verbatim}
\end{mycode}

\hspace{1in}\includegraphics[scale=.9, bb = 0 -1 1 -1]{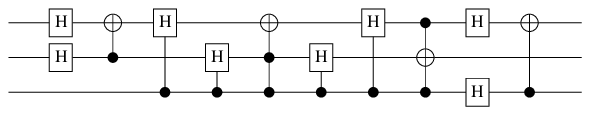}
\end{minipage}

\noindent Another block structure operator provided by Quipper is
\begin{mycode}
\begin{verbatim}
with_ancilla :: (Qubit -> Circ a) -> Circ a.
\end{verbatim}
\end{mycode}
This operator can be used to provide an ancilla qubit (temporary
scratch space) to a block of gates. The ancilla is initially in state
$\ket0$, and the code is expected to return it to state $\ket0$ at the
end of the block. The following example also illustrates the use of
the {\tt controlled} operator, which is an infix version of {\tt
  with\_controls}. The controls are specified to the right of the 
operator, and can be a tuple of qubits.

\noindent\begin{minipage}{\columnwidth}
\begin{mycode}
\begin{verbatim}
mycirc3 :: Qubit -> Qubit -> Qubit 
  -> Circ (Qubit, Qubit, Qubit)
mycirc3 a b c = do
  with_ancilla $ \x -> do
    qnot x `controlled` (a,b)
    hadamard c `controlled` x
    qnot x `controlled` (a,b)
  return (a,b,c)
\end{verbatim}
\end{mycode}

\hspace{1.8in}\includegraphics[scale=1.1, bb=0 -1 1 -1]{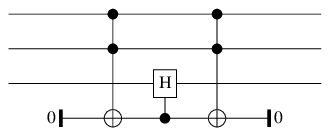}
\end{minipage}

\subsubsection{Circuit operators}

In addition to the gate-by-gate circuit construction paradigm, Quipper
also provides powerful higher-order operators that operate on entire
quantum functions. The block-structuring commands of the previous
subsection are examples of simple higher-order operators. Other
high-level operators provided by Quipper include operators for
reversing, iterating, and transforming quantum procedures, as well as
a general mechanism for turning classical boolean procedures into
quantum oracles.

The {\tt reverse\_simple} operator takes a quantum function and
returns its inverse:

\noindent\begin{minipage}{\columnwidth}
\begin{mycode}
\begin{verbatim}
timestep :: Qubit -> Qubit -> Qubit 
  -> Circ (Qubit, Qubit, Qubit)
timestep a b c = do
    mycirc a b
    qnot c `controlled` (a,b)
    reverse_simple mycirc (a,b)
    return (a,b,c)
\end{verbatim}
\end{mycode}

\hspace{1.8in}\includegraphics[scale=1.1, bb = 0 -10 1 -10]{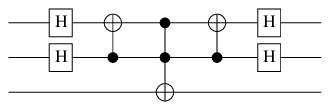}
\end{minipage}

It is important to realize that reversing a circuit is not necessarily
an operation to be performed just on the output of a program (say, by
a separate tool). Many quantum algorithms require a circuit to be
reversed in the middle of a computation, perhaps within a nested
subroutine.

The operator {\tt decompose\_generic} decomposes a quantum circuit
into a specified set of elementary gates. The inputs and outputs of
the circuit are unchanged, so the resulting quantum circuit has the 
same type as the original circuit. 
The decomposition is achieved by first decomposing 
multiply-controlled gates into Toffoli gates, and then decomposing 
the Toffoli gates into binary gates 
{\cite[Sec.~4.3]{Nielsen-Chuang-2002}}. For example, the following
decomposes the circuit from the previous example into binary gates:

\noindent\begin{minipage}{\columnwidth}
\begin{mycode}
\begin{verbatim}
timestep2 :: Qubit -> Qubit -> Qubit 
  -> Circ (Qubit, Qubit, Qubit)
timestep2 = decompose_generic Binary timestep
\end{verbatim}
\end{mycode}
  \begin{center}
    \includegraphics[scale=1]{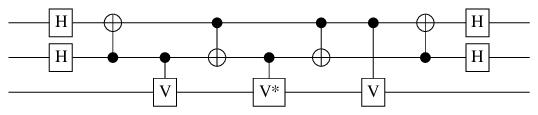}
  \end{center}
\end{minipage}

\subsubsection{Boxed subcircuits}

Quipper circuits can be very large; for example, in
Section~\ref{sec:triangle}, we use Quipper to describe a circuit of
over 30 trillion gates. In order to be able to store and manipulate
such large circuits efficiently, Quipper provides a feature called
{\em hierarchical circuits} or {\em boxed subcircuits}. The idea is
simple: if a certain subcircuit is used multiple times throughout a
larger circuit, the programmer has the option to ``box'' it. In this
case, the subcircuit will be replaced by a single named gate, with a
separate definition on the side. Boxed subcircuits can be nested,
leading to a hierarchy of circuits. The Quipper operator for
introducing a boxed subcircuit is called {\tt box}. It takes a name
and a circuit-generating function as its arguments. See
Section~\ref{sec:triangle} for examples.

\subsubsection{Run functions}

As we have already seen, in Quipper, the {\em description} of circuits
is separated from {\em what to do} with them. Thus, the same
subroutine can be used, for example, to run a circuit on a quantum
device, or to construct and manipulate it in memory. We believe that
this separation provides a useful abstraction to programmers.

What to do with a circuit is determined by different {\em run
  functions} for the {\tt Circ} monad. For example, the function {\tt
  print\_gen\-eric} can be used to print a circuit in a number of
available output formats (such as text, PostScript, and PDF). Quipper
also provides a function {\tt run\_generic} to simulate a circuit
(this is necessarily inefficient on a classical computer). The more
specialized functions {\tt run\_classical\_generic} and {\tt
  run\_clifford\_generic} can be used to simulate certain classes of
circuits efficiently; this is especially useful in testing oracles.

\subsection{Quipper's extensible quantum data types}

Following the strategy first presented in Altenkirch and Green's work
on the Quantum IO monad~\cite{Altenkirch-Green-2009}, Quipper uses
Haskell's type classes to provide an abstract view of the notion of
quantum data. A type class can be thought of as a property that a type
may satisfy; the property comes with a set of functions. The strength
of type classes is that they can be defined by induction on the
structure of types.

In Quipper, the notion of quantum data is represented by the type
class {\tt QCData}. The most basic members of this type class are {\tt
  Qubit} and {\tt Bit}, representing a quantum bit and a classical bit
in a circuit, respectively. Expanding on this, tuples of quantum data
are quantum data, lists of quantum data are quantum data, and so
forth:

\begin{mycode}
\begin{verbatim}
instance (QCData a, QCData b) => QCData (a,b) where ...
instance (QCData a)           => QCData [a]   where ...
\end{verbatim}
\end{mycode}

Quipper also comes with a number of libraries defining additional
kinds of quantum data. For example, there is an arithmetic library
that defines {\tt QDInt}, a type of fixed-size signed quantum
integers, and a real number library defining a type {\tt FPReal} of
fixed-size, fixed-point real numbers.

Certain generic quantum operations can be defined at any {\tt QCData}
instance, rather than just qubits. For example, the built-in Quipper
function {\tt controlled\_not}, which applies a controlled not 
operation to each corresponding pair of qubits from two quantum data 
structures, has type:

\begin{mycode}
\begin{verbatim}
controlled_not :: (QCData q) => q -> q -> Circ (q, q).
\end{verbatim}
\end{mycode}

Quipper also provides a type class {\tt QShape}, which takes 3
arguments and represents the relationship between the quantum input,
classical input, and classical parameter versions of a type, as
described in Section~\ref{sssec:parameter-input}. For example, we have
\begin{mycode}
\begin{verbatim}
instance QShape Bool Qubit Bit
instance (QShape b q c, QShape b' q' c') 
  => QShape (b,b') (q,q') (c,c')
instance QShape IntM QDInt CInt
\end{verbatim}
\end{mycode}

Most of Quipper's built-in circuit generating functions natively use
these representations. For example, the functions for initialization
and measurement of quantum data have the type
\begin{mycode}
\begin{verbatim}
qinit :: QShape b q c => b -> Circ q
measure :: QShape b q c => q -> Circ c
\end{verbatim}
\end{mycode}
For example, we can use {\tt qinit} to create a pair of quantum bits:
\begin{mycode}
\begin{verbatim}
example = do
   (p,q) <- qinit (False,False)
   ...
\end{verbatim}
\end{mycode}

\subsection{Oracles in Quipper}
\label{ssec:quipperoracle}

Although appending gates to quantum circuits is an important part of
many quantum algorithms, the most challenging part for the quantum
programmer --- and the biggest, in terms of number of gates produced
--- is often the implementation of classical oracles. Such oracles are
boolean functions represented as reversible quantum circuits. They
are problem specific and can be quite complicated. For example, Shor's
factoring algorithm {\cite{shor94}} relies on an oracle for computing the
modular exponentiation $f(x) = a^x\ (\mymod N)$, where $N$ is the
integer to be factored. In the Triangle Finding algorithm, described
in more detail in Section~\ref{sec:triangle} below, an oracle is used
to define the edges of the graph that is the input to the algorithm.

Quipper provides powerful facilities for programming oracles in a
natural way.

\subsubsection{Automatic generation of quantum oracles}

The implementation of a quantum oracle ``by hand'' usually requires
four separate steps. The first step is to express the oracle as a
classical program acting on classical data types. The second step is
to translate this program to a classical circuit for the given input
size. The third step is to change the classical circuit to a quantum
circuit, possibly introducing many ancillas to hold intermediate or
``scratch space'' values. The fourth step is to make this quantum
circuit reversible, using the standard trick of replacing the function
$x\mapsto f(x)$ by a reversible function $(x,y)\mapsto(x,y\oplus
f(x))$, while also uncomputing any scratch space used by the function
$f$.

In Quipper, all of these steps but the first one can be automated.
Consider, for example, a very simple oracle that inputs a list of
booleans and outputs their parity (even or odd). This can be naturally
expressed as a functional program:
\begin{mycode}
\begin{verbatim}
build_circuit
f :: [Bool] -> Bool
f as = case as of
  [] -> False
  [h] -> h
  h:t -> h `bool_xor` f t
\end{verbatim}
\end{mycode}
The keyword {\tt build\_circuit} is built into Quipper (incidentally,
it has been implemented in a very interesting way, using a custom
pre-processor and Template Haskell~\cite{TH}). Its purpose is to
perform an operation that we call {\em circuit lifting}, automating
steps 2 and 3 above. Specifically, the effect of the {\tt
  build\_circuit} keyword is to produce, at compile time, a
circuit-generating function {\tt template\_f} in addition to the
function {\tt f}. The type of the function {\tt template\_f} is
obscure, but can be made useful by passing it through Quipper's {\tt
  unpack} operation:
\begin{mycode}
\begin{verbatim}
unpack template_f :: [Qubit] -> Circ Qubit
\end{verbatim}
\end{mycode}
The function {\tt template\_f} automatically produces a circuit
computing the same operation as {\tt f}. For example, when
applied to a list of 4 qubits, it produces:
\begin{center}
  \includegraphics[scale=0.53]{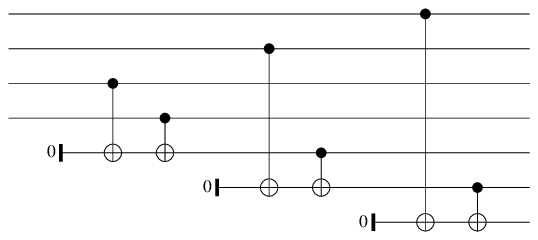}
\end{center}
Note how the top four qubits are the inputs, the bottom qubit is the
output, and the remaining two qubits are scratch space. Finally, the
fourth step, to make the circuit reversible and uncompute the scratch
space, is taken care of by the Quipper operator {\tt
  classical\_to\_reversible}:
\begin{mycode}
\begin{verbatim}
classical_to_reversible :: (Datable a, QCData b) => 
                (a -> Circ b) -> (a,b) -> Circ (a,b)
\end{verbatim}
\end{mycode}
For example, here is the circuit produced by 
\begin{mycode}
\begin{verbatim}
classical_to_reversible (unpack template_f):
\end{verbatim}
\end{mycode}
\begin{center}
  \includegraphics[scale=0.53]{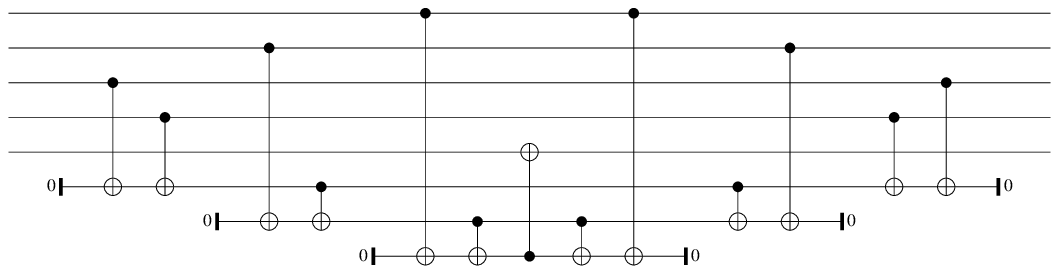}
\end{center}
Note that in this circuit, the top four qubits are inputs, the fifth
qubit is the output, and all intermediate ancillas have been
uncomputed.

Quipper's circuit lifting operation is extremely versatile. We have
used it to implement oracles containing millions of gates. For
example, our implementation of the Boolean Formula algorithm uses an
oracle that determines the winner for a given final position in the
game of Hex. It uses a flood-fill algorithm, which we implemented as a
functional program and converted to a circuit using the circuit
lifting operation. The resulting oracle consists of 2.8 million
gates. Similarly, our implementation of the Linear Systems algorithm
makes liberal use of arithmetic and analytic functions, such as
$\sin(x)$ and $\cos(x)$, which were implemented using the circuit
lifting feature. The circuit created for $\sin(x)$, over a 
32$+$32 qubit fixed-point argument, uses 3273010 gates.

\section{The Triangle Finding algorithm in Quipper}\label{sec:triangle}

We give some details of our implementation of the Triangle Finding
algorithm in Quipper.

\subsection{Background}

An instance of the \emph{Triangle Finding problem} {\cite{Ma_Sa_Sz1,Ch_Ko1}} is given by an undirected simple graph $G$ containing exactly one triangle $\Delta$. The graph is given by an oracle function $f$, such that, for any two nodes $v,w$ of $G$, $f(v,w)=1$ if $(v,w)$ is an edge of $G$ and $f(v,w)=0$ otherwise. To solve an instance of the Triangle Finding problem is to find the set of vertices $\{e_1,e_2,e_3\}$ forming $\Delta$ by querying $f$.

The \emph{Triangle Finding algorithm}, as described in \cite{Ma_Sa_Sz1} and \cite{Ch_Ko1}, works by performing a Grover-based quantum walk on a larger graph $H$, called the {\em Hamming graph} associated to $G$. It is designed to find $\Delta$ with high probability. The algorithm is parametric on an oracle defining the graph $G$. In our implementation, the oracle is a changeable part, but we have implemented a particular pre-defined oracle specified by the QCS program. This oracle injects $G$ into the space $\{0,1,\ldots,2^l-1 \}$ of $l$-bit integers, and each oracle call requires the extensive use of modular arithmetic.

The overall algorithm is parameterized on integers $l,n$ and $r$ specifying respectively the length $l$ of the integers used by the oracle, the number $2^n$ of nodes of $G$ and the size $2^r$ of Hamming graph tuples.

\subsection{Top-level structure}

The Quipper implementation of the Triangle Finding algorithm is broken down into six modules:
\begin{compactitem}
  \item \verb!Definitions!: global definitions used throughout the algorithm.
  \item \verb!QWTFP!: the quantum walk algorithm and its subroutines.
  \item \verb!Oracle!: the oracle and its subroutines.
  \item \verb!Main!: a command line interface.
  \item \verb!Simulate!: a test suite for the oracle.
  \item \verb!Alternatives!: alternatives and/or generalization of certain algorithms.
\end{compactitem}
These can be compiled into an executable program {\tt tf}. Its command line interface allows the user, for example, to plug in different oracles, show different parts of the circuit, select a gate base, select different output formats, and select parameter values for $l$, $n$ and $r$. Some usage examples are provided throughout the remainder of this section as we discuss our implementation. 

\subsection{Code samples}

The quantum walk part of the algorithm is broken into about 20 subroutines, and the oracle consists of 8 subroutines. For brevity, we only present the code for one of each: \verb!o4_POW17! and \verb!a6_QWSH!. Although relatively simple, these subroutines are good illustrations of some of Quipper's key features.

\subsubsection{The subroutine {\tt o4\_POW17}}

\begin{figure}
\begin{center}
\includegraphics[width=\columnwidth]{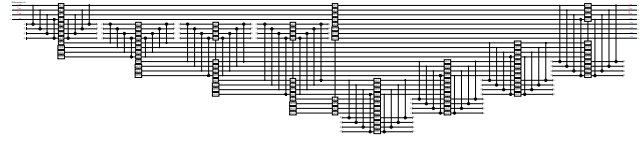}
\caption{The circuit for {\tt o4\_POW17}}\label{fig-o4-pow17}
\end{center}
\end{figure}

\begin{figure*}
\begin{center}
\includegraphics[width=\textwidth]{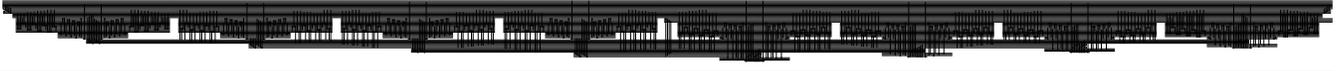}
\caption{The circuit for {\tt o8\_MUL}}\label{fig-o8-mul}
\end{center}
\end{figure*}

The subroutine \verb!o4_POW17! is an arithmetic function used by the oracle. It computes the seventeenth power of a quantum integer and stores the result in a fresh integer register. It proceeds by first raising its input $x$ to the 16th power by repeated use of a squaring subroutine, and then multiplies $x$ and $x^{16}$ to get the desired result. In the following Quipper code, \verb!QIntTF! denotes the type of quantum integers used by the oracle, which happen to be $l$-bit integers with arithmetic taken modulo $2^l-1$ (not $2^l)$:

{\footnotesize
\begin{mycode}
\begin{verbatim}
o4_POW17 :: QIntTF -> Circ (QIntTF,QIntTF)
o4_POW17 = box "o4" $ \x -> do
 comment_with_label "ENTER: o4_POW17" x "x"
  
 (x, x17) <- with_computed_fun x
  (\x -> do
   (x,x2) <- square x
   (x2,x4) <- square x2
   (x4,x8) <- square x4      
   (x8,x16) <- square x8
   return (x,x2,x4,x8,x16))
    
  (\(x,x2,x4,x8,x16) -> do
   (x,x16,x17) <- o8_MUL x x16
   return ((x,x2,x4,x8,x16),x17))
      
 comment_with_label "EXIT: o4_POW17" (x,x17) ("x","x17")
 return (x, x17)
\end{verbatim}
\end{mycode}
}   

We note the use of the pre-defined Quipper operators \verb!box!, {\tt com\-ment\_with\_label} and \verb!with_computed_fun!. The operator \verb!box! introduces a boxed subcircuit. The operator \verb!comment_with_label! inserts a comment and some qubit labels in the generated circuits. Such comments have proven to be quite useful in reading large circuits. The operator \verb!with_computed_fun! automates the reversing of intermediary computations: the first block of code (in this case, applications of \verb!square! producing $x^2, x^4, x^8$ and $x^{16}$) is reversed once the second block of code (here \verb!o8_MUL!) has been applied. Because the uncomputation of intermediate results is such a common operation in quantum computing, the use of operators like \verb!with_computed_fun! helps to avoid unnecessary and error-prone code repetitions. All three of these Quipper features can be seen in the circuit for \verb!o4_POW17! with parameter values $l=4,n=3$ and $r=2$ shown in Figure~\ref{fig-o4-pow17}. This circuit is produced by the command line \verb!./tf -s pow17 -l 4 -n 3 -r 2!.

We note that some of the circuits shown here have too many gates to be
legible in a printed version of this paper; however, in the PDF
version, it is possible to zoom in to see individual gates.

In the circuit in Figure~\ref{fig-o4-pow17}, the vertical strings of squares marked {\tt o8} represent invocations of a boxed subcircuit. Each of them denotes an invocation of the subroutine \verb!o8_MUL! for multiplication, or its inverse. The full definition of \verb!o8_MUL! is shown in Figure~\ref{fig-o8-mul}. 

It is possible to inline the boxed subcircuits within \verb!o4_POW17!, but the resulting circuit would be too large to be usefully included here. However, we can use Quipper's gate counting feature to provide some statistics about this circuit. The is done via the command line option \verb!-f gatecount!. It will compute a gate count for each boxed subcircuit called by \verb!o4_POW17!, together with an aggregated gate count for the circuit with all boxed subcircuits inlined. For $l=4$, $n=3$, $r=2$, the aggregated gate count for \verb!o4_POW17! is:

{\footnotesize
\begin{verbatim}
Aggregated gate count:
 1636: "Init0"
 3484: "Not", controls 1
  288: "Not" controls 1+1
 2592: "Not", controls 2
 1632: "Term0"
Total gates: 9632
Inputs: 4
Outputs: 8
Qubits in circuit: 71
\end{verbatim}
}   

In words, this circuit has 4 inputs, 8 outputs, and uses a total of 71
qubits (including ancillas) and 9632 elementary gates. Of these gates,
about one third are qubit initializations and terminations, and the
remainder are controlled-not gates with 1 or 2 controls.
In gate counts provided by Quipper a distinction is made between positive and negative controls. If a gate $G$ has $a$ positive controls (``filled dots'') and $b$ negative controls (``empty dots''), the gate count will read: \verb!"G", controls a+b!. Moreover, \verb!a+0! is written \verb!a!.

\subsubsection{The subroutine {\tt a6\_QWSH}}

The subroutine \verb!a6_QWSH! implements a walk step on the Hamming graph. By definition, the nodes of the Hamming graph associated to $G$ are tuples of nodes of $G$, such that two such tuples are adjacent if they differ in exactly one coordinate. \verb!a6_QWSH! proceeds in two steps. In the first step, it arbitrarily chooses an index $i$ and a node $v$ of $G$. In the second step, it replaces a Hamming tuple $T$ by an adjacent one $T'$ by swapping the $i$-th component of $T$ with $v$, and updates the register containing the edge information concerning nodes in $T'$. The corresponding Quipper code is the following:

{\footnotesize
\begin{mycode}
\begin{verbatim}
a6_QWSH :: QWTFP_spec -> (IntMap QNode) -> QDInt 
 -> QNode -> (IntMap (IntMap Qubit))
  -> Circ (IntMap QNode, QDInt, QNode, 
            IntMap (IntMap Qubit))
a6_QWSH oracle@(n,r,edgeOracle,qram) = 
  box "a6" $ \tt i v ee -> do
  comment_with_label "ENTER: a6_QWSH" 
    (tt, i, v, ee) ("tt", "i", "v", "ee")
  with_ancilla_init (replicate n False) $ \ttd -> do 
    with_ancilla_init (intMap_replicate (2^r) False) $ 
     \eed -> do
      (i,v) <- a7_DIFFUSE (i,v)
      ((tt,i,v,ee,ttd,eed),_) <- 
\end{verbatim}
\end{mycode}
}{\footnotesize
\begin{mycode}
\begin{verbatim}
       with_computed_fun (tt,i,v,ee,ttd,eed)
        (\(tt,i,v,ee,ttd,eed) -> do
          (i,tt,ttd) <- qram_fetch qram i tt ttd 
          (i,ee,eed) <- a12_FetchStoreE i ee eed
          (tt,ttd,eed) <- a13_UPDATE oracle tt ttd eed
          (i,tt,ttd) <- qram_store qram i tt ttd
          return (tt,i,v,ee,ttd,eed))
            
        (\(tt,i,v,ee,ttd,eed) -> do
          (ttd,v) <- a14_SWAP ttd v
          return ((tt,i,v,ee,ttd,eed),()))
  
      comment_with_label "EXIT: a6_QWSH" 
       (tt, i, v, ee) ("tt", "i", "v", "ee")
      return (tt,i,v,ee)
\end{verbatim}
\end{mycode}
}   

Here, the Quipper operator \verb!with_ancilla_init! creates a list of $n$ ancillas, whose scope is restricted to a local block of code. The circuit for \verb!a6_QWSH! with parameter values $l=4, n=3$ and $r=2$ is:

\begin{center}
\includegraphics[scale=0.6]{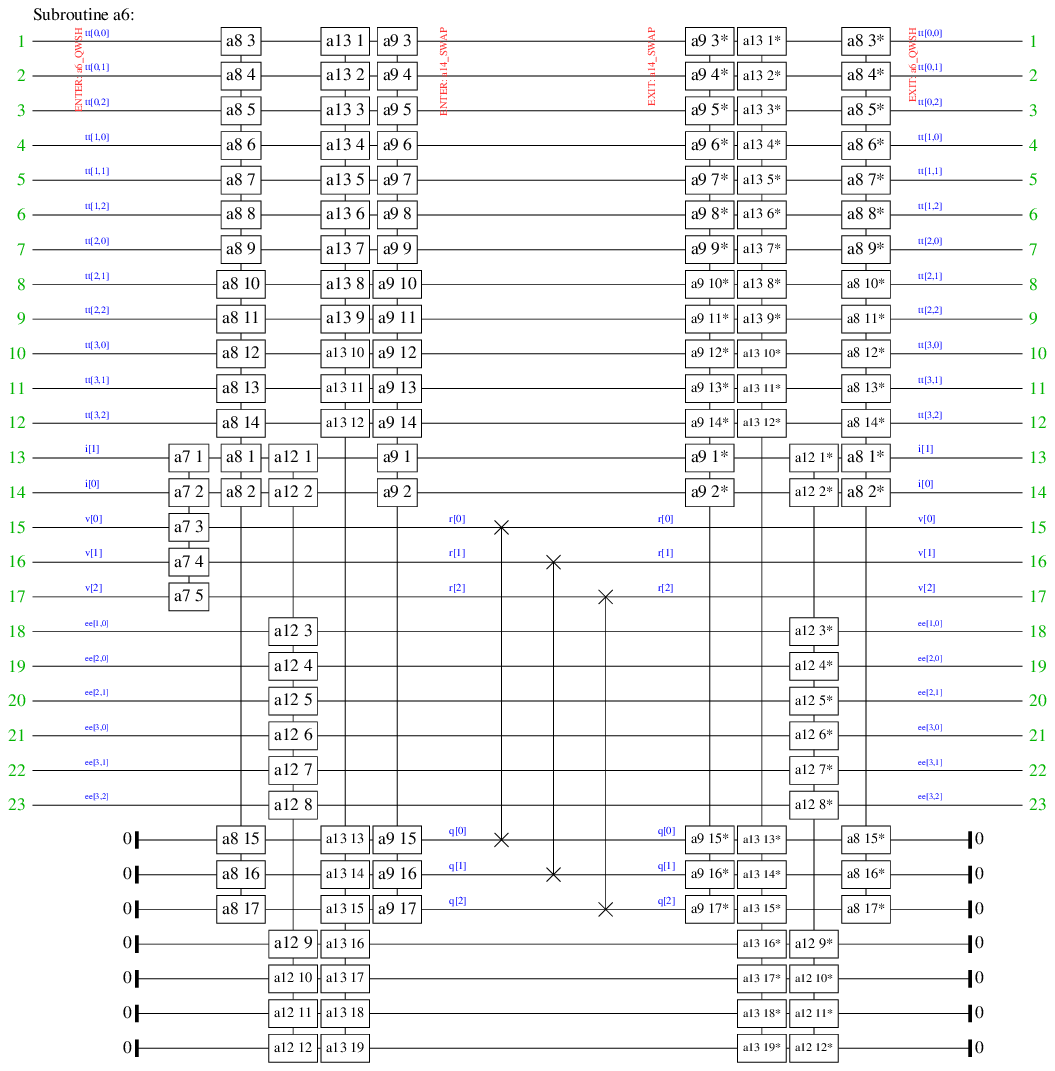}
\end{center}

In this circuit, the first boxed subcircuit corresponds to the diffusion of the index $i$ and node $v$. The remaining boxed subcircuits denote the qRam operations before and after the node swap. 

\subsection{Aggregate gate counts}

The command line 
\begin{center}
  \verb!./tf -f gatecount -O -o orthodox -l 31 -n 15 -r 9!
\end{center}
computes the gate count for just the oracle, with parameter values $n=15$, $l=31$ and $r=9$. It counts 2051926 total gates and 1462 qubits.
The command line 
\begin{center}
  \verb!./tf -f gatecount -o orthodox -l 31 -n 15 -r 6!
\end{center}
produces the gate counts for the complete algorithm, including repeated quantum walk steps with inlined oracle invocations. On a standard laptop, this runs to completion in under two minutes and produces a count of 30189977982990 (over 30 trillion) total gates and 4676 qubits. 

\section{Comparing Quipper and QCL}\label{sec:qcl}

To enable a direct comparison between Quipper and QCL, we implemented
identical versions of the Binary Welded Tree algorithm {\cite{BWT}} in
both programming languages, using a hand-coded oracle. For further
comparison, we also gave a second implementation of an equivalent
oracle, using Quipper's {\tt build\_circuit} mechanism to
automatically generate the (non-optimized) oracle from classical
functional code as explained in Section~\ref{ssec:oracles}. We
generated the main circuit for the BWT algorithm for each of the three
different implementations, using the same parameters in each case. The
results are summarized in the following table.

\begin{center}
\scriptsize
\begin{tabular}{l|c|c|c}
 \bf & \bf QCL ``direct'' & \bf Quipper ``orthodox'' & \bf Quipper
 ``template'' \\\hline
 Init & 58 & 313 & 777 \\
 Not & 746 & 8 & 0 \\
 CNot$_1$ &  9012 & 472 & 344 \\
 CNot$_2$ & 7548 & 768 & 1760 \\
 $e^{-itZ}$ & 4 & 4 & 4 \\
 $W$  & 48 & 48 & 48 \\
 Term & 0 & 307 & 771 \\
 Meas & 0 & 6 & 6 \\\hline
 Total & 17358 & 1300 & 2156 \\\hline
 Qubits & 58 & 26 & 108 \\
\end{tabular}
\end{center}

Here ``Init'', ``Term'', and ``Meas'' refer to Quipper's qubit
initialization, termination, and measurement gates. These are not
directly comparable between QCL and Quipper, because Quipper
explicitly tracks the scope of ancillas whereas QCL does
not. ``Total'' refers to the total number of logical gates excluding
initialization, termination, and measurement. ``Qubits'' refers to the
total number of qubits used in each circuit, i.e., the height of the
circuit.

It is apparent that the QCL code produces far more gates than its
Quipper counterpart, even when the hand-coded oracle in QCL is
compared to the automatically generated oracle in Quipper. Moreover,
the QCL circuit uses twice as many qubits as the Quipper version with
the same oracle.  On the other hand, the Quipper implementation with
automatically generated oracle uses more ancillas than QCL, but does
so with fewer gates.

\section{Conclusion}

We have presented Quipper, a scalable functional quantum programming
language. We demonstrated its usability by implementing seven
non-trivial quantum algorithms, chosen to represent a broad range of
quantum computing capabilities. The algorithms were implemented by a
team of 11 geographically distributed Quipper programmers. Programming
the seven algorithms required approximately 55 man months and resulted
in a representation usable for resource estimation using realistic
problem sizes. On this basis we conclude that Quipper is both usable
and useful.

One of the issues left for future work in Quipper is the improvement
of compile-time type checking. Thanks to its Haskell implementation,
Quipper already catches many ordinary type errors at compile time.
However, in the absence of a linear type system, certain properties,
such as non-duplication of quantum data, must be checked at
runtime. Developing a fully-featured type system is the next step in
Quipper's development, and is a work in progress.

\section{Acknowledgements}
\label{sec:conclusion}

Thanks to Jonathan M. Smith for his helpful comments.

Supported by the Intelligence Advanced Research Projects Activity
(IARPA) via Department of Interior National Business Center contract
number D11PC20168. The U.S. Government is authorized to reproduce
and distribute reprints for Governmental purposes notwithstanding any
copyright annotation thereon. Disclaimer: The views and conclusions
contained herein are those of the authors and should not be
interpreted as necessarily representing the official policies or
endorsements, either expressed or implied, of IARPA, DoI/NBC, or the
U.S. Government. Supported by NSERC.

\bibliographystyle{abbrv}
\bibliography{paper}

\begin{thebibliography}{10}

\bibitem{Altenkirch-Green-2009}
T.~Altenkirch and A.~S. Green.
\newblock The {Quantum IO Monad}.
\newblock In S.~Gay and I.~Mackie, editors, {\em Semantic Techniques in Quantum
  Computation}, pages 173--205. Cambridge University Press, 2009.

\bibitem{BF}
A.~Ambainis, A.~M. Childs, B.~Reichardt, R.~{\v S}palek, and S.~Zhang.
\newblock Any {AND}-{OR} formula of size $n$ can be evaluated in time
  $n^{\frac12+o(1)}$ on a quantum computer.
\newblock {\em SIAM J. Comput.}, 39:2513–--2530, 2010.

\bibitem{Ch_Ko1}
A.~Childs and R.~Kothari.
\newblock Quantum query complexity of minor-closed graph properties.
\newblock In {\em Proceedings of the 28th Symposium on Theoretical Aspects of
  Computer Science}, pages 661--672, 2011.

\bibitem{BWT}
A.~M. Childs, R.~Cleve, E.~Deotto, E.~Farhi, S.~Gutmann, and D.~A. Spielman.
\newblock Exponential algorithmic speedup by a quantum walk.
\newblock In {\em Proceedings of the Thirty-Fifth Annual ACM Symposium on
  Theory of Computing}, pages 59--68, 2003.

\bibitem{Claessen-2001}
K.~Claessen.
\newblock {\em Embedded Languages for Describing and Verifying Hardware}.
\newblock PhD thesis, Chalmers University of Technology and G{\"o}teborg
  University, 2001.

\bibitem{Deutsch-1985}
D.~Deutsch.
\newblock Quantum theory, the {Church-Turing} principle and the universal
  quantum computer.
\newblock {\em Proceedings of the Royal Society of London, Series A},
  400(1818):97--117, 1985.

\bibitem{Gay-2006}
S.~J. Gay.
\newblock Quantum programming languages: Survey and bibliography.
\newblock {\em Mathematical Structures in Computer Science}, 16(04):581--600,
  2006.

\bibitem{CN}
S.~Hallgren.
\newblock Polynomial-time quantum algorithms for {P}ell's equation and the
  principal ideal problem.
\newblock {\em J. ACM}, 54(1):4:1--4:19, Mar. 2007.

\bibitem{LS}
A.~W. Harrow, A.~Hassidim, and S.~Lloyd.
\newblock Quantum algorithm for linear systems of equations.
\newblock {\em Phys. Rev. Lett.}, 103(15):150502, 2009.

\bibitem{BAA}
{IARPA Quantum Computer Science Program}.
\newblock {Broad Agency Announcement IARPA-BAA-10-02}.
\newblock Available from
  https://www.fbo.gov/notices/637e87ac1274d030ce2ab69339ccf93c, April 2010.

\bibitem{jordan-qzoo}
S.~Jordan.
\newblock {\tt http://math.nist.gov/quantum/zoo/}.
\newblock Electronic resource.

\bibitem{Knill-1996}
E.~H. Knill.
\newblock Conventions for quantum pseudocode.
\newblock LANL report LAUR-96-2724, 1996.

\bibitem{TF}
F.~Magniez, M.~Santha, and M.~Szegedy.
\newblock Quantum algorithms for the triangle problem.
\newblock quant-ph/0310134, 2003.

\bibitem{Ma_Sa_Sz1}
F.~Magniez, M.~Santha, and M.~Szegedy.
\newblock Quantum algorithms for the triangle problem.
\newblock In {\em Proceedings of the 16th annual ACM-SIAM symposium on Discrete
  algorithms}, pages 1109--1117, 2005.

\bibitem{Nielsen-Chuang-2002}
M.~A. Nielsen and I.~L. Chuang.
\newblock {\em Quantum Computation and Quantum Information}.
\newblock Cambridge University Press, 2002.

\bibitem{Omer-2000}
B.~\"{O}mer.
\newblock Quantum programming in {QCL}.
\newblock Master's thesis, Institute of Information Systems, Technical
  University of Vienna, 2000.

\bibitem{SV}
O.~Regev.
\newblock Quantum computation and lattice problems.
\newblock {\em SIAM J. Comput.}, 33(3):738--760, 2004.

\bibitem{Selinger-Valiron-2006}
P.~Selinger and B.~Valiron.
\newblock A lambda calculus for quantum computation with classical control.
\newblock {\em Mathematical Structures in Computer Science}, 16(3):527--552,
  2006.

\bibitem{Selinger-Valiron-2009}
P.~Selinger and B.~Valiron.
\newblock Quantum lambda calculus.
\newblock In S.~Gay and I.~Mackie, editors, {\em Semantic Techniques in Quantum
  Computation}, pages 135--172. Cambridge University Press, 2009.

\bibitem{TH}
T.~Sheard and S.~Peyton~Jones.
\newblock Template metaprogramming for {Haskell}.
\newblock In {\em Proc. Haskell Workshop}, 2002.

\bibitem{shor94}
P.~Shor.
\newblock Algorithms for quantum computation: discrete logarithms and
  factoring.
\newblock In {\em Proceedings, 35th Annual Symposium on Foundations of Computer
  Science}. CA: IEEE Press, 1994.

\bibitem{Tonder-2004}
A.~van Tonder.
\newblock A lambda calculus for quantum computation.
\newblock {\em SIAM Journal of Computing}, 33(5):1109--1135, 2004.

\bibitem{GSE}
J.~D. Whitfield, J.~Biamonte, and A.~Aspuru-Guzik.
\newblock Simulation of electronic structure {Hamiltonians} using quantum
  computers.
\newblock {\em Molecular Physics}, 109(5):735--750, 2011.

\end{thebibliography}

\end{document}